\documentclass[pre,twocolumn,a4paper,showpacs,superscriptaddress]{revtex4}
\usepackage{amsmath}
\usepackage{amsfonts}
\usepackage{graphicx}
\begin{document}
\title{Dynamical evolution  of  quantum oscillators towards equilibrium}
\author{A. R. Usha Devi}
\email{arutth@rediffmail.com}
\affiliation{Department of Physics, Bangalore University, 
Bangalore-560 056, India}
\affiliation{H. H. Wills Physics Laboratory, University of Bristol, Bristol BS8 1TL, UK}
\affiliation{Inspire Institute Inc., McLean, VA 22101, USA.}
\author{A. K. Rajagopal}
\affiliation{Inspire Institute Inc., McLean, VA 22101, USA.}
\date{\today}

\begin{abstract}
A pure quantum state of large number $N$ of oscillators, interacting via harmonic coupling, evolves such that 
any small subsystem $n<<N$ of the global state approaches equilibrium. This provides a novel example where 
stationarity emerges as a natural phenomena under quantum dynamics alone, 
with no necessity to bring in any additional statistical postulates.  Mixedness of equilibrated subsystems 
consisting of $1, 2, \ldots, n<<N$ clearly indicates that small subsystems are  entangled with the rest of the 
state i.e., the bath.  Every  single mode oscillator  is found to relax in a mixed density matrix of the 
Boltzmann canonical form. In two oscillator stationary subsystems, intra-entanglement within the `system' 
oscillators is found to exist  when the magnitude of the  squeezing parameter of the bath is comparable in 
magnitude with that of the coupling strength. 
\end{abstract}

\maketitle
\section{Introduction}
Deducing the statistical distribution in many particle systems as an intrinsic property, resulting solely from 
quantum dynamics, has attracted much attention in the literature~\cite{many}. In this context, considerable 
interest has been evoked recently~\cite{Tasaki, PR, SP} on a deeper understanding of the basic mechanism of 
equilibration~\cite{fn0}, occurring  as a natural consequence of quantum dynamical evolution - without invoking 
any additional statistical assumptions.  More specifically, equilibration is realized entirely in a quantum 
mechanical setting -- with the key element ascribed to quantum entanglement between the system and the 
environment. 
Erstwhile statistical postulates on ensemble averaging over initial distributions are not 
required at all -- as quantum dynamics of individual pure states of a many body physical system itself leads to 
equilibration of smaller subsystems. This features an exciting foundational development, where 'subjective' lack 
of knowledge in terms of statistical ensemble averaging   is replaced by the 'objective' randomness  due to 
entanglement~\cite{fn1}.   

Based on  powerful general arguments, Linden et. al. established~\cite{SP}  that an overwhelming majority of 
pure quantum states of interacting large quantum systems evolve such that any small subsystem approaches a 
stationary state. This brings out an important implication: dynamics of almost every pure many body quantum 
state, envisages stationarity of any small subsystem as an inherent property, with all the statistical 
ingredients already built within the basic quantum framework itself.   
 
In this paper, we present an explicit analysis of the quantum evolution of an  initially uncoupled pure squeezed 
state of a large number $N$ of  oscillators, subjected to a harmonic interaction Hamiltonian, resulting 
eventually in equilibrium of any small subsystem of $n<<N$ oscillators of the global pure state -  with the rest 
acting as  the bath. Note that in Ref.~\cite{SP}, the {\em smallness} of the system in 
relation to the size of the bath is described by their respective dimensions. In the  present case, individual 
systems constituting the whole state are infinite dimensional (being harmonic oscillators) and so,   
 the {\em smallness} of the subsystems of the global quantum system consisting of $N$ oscillators 
 is  expressed legitimately in terms of the number  $n<<N$ of a subset of oscillators 
 under consideration.  Essentially, we find that every small subsystem of oscillators 
 tend to relax 
in a stationary state,  specified by a mixed density matrix 
- which is indeed a signature of quantum entanglement of the `system' and the `bath', emerging due to quantum 
dynamics. It may be worth pointing out here that due to its mathematical transparency and simplicity, the 
physical  model of a linear assembly of coupled oscillators  has 
played a paradigmatic role  in understanding difficult formal principles underlying statistical 
mechanics~\cite{KP}.  A detailed investigation of quantum dynamics of pure states leading to equilibrated small 
subsystems in this model would therefore be illuminating. 

The paper is organized as follows: In Sec.~II, we describe the physical model of $N$ harmonically coupled  
oscillators and discuss its exact solution using symplectic transformations -- which offers a most natural  
elegant approach to the problem of interest. We then identify the $N\times N$ symplectic transformation 
corresponding to unitary time evolution in this model. This is followed by Sec.~III, where we analyze the 
time evolution of a global pure uncoupled squeezed state (which is not an eigenstate of the Hamiltonian) of the 
system-bath oscillators.  Under the assumptions of continuum limit $N\rightarrow \infty$ and  weak coupling 
approximation we show that any small subsystem of the whole pure state of $N$ oscillators exhibits a stationary 
long time behavior. Every single mode oscillator system is shown to relax in a mixed state of Boltzmann  
canonical form, with effective temperature related to the squeezing parameter of the bath oscillators.It is also 
shown that any two oscillator system approach a stationary mixed state with an intra-entanglement surviving    
whenever the magnitude of the bath squeezing parameter is comparable in magnitude with that of the coupling 
strength. Sec.~IV has concluding remarks.              
\section{The physical model}
 We consider a long chain of coupled harmonic oscillators, the Hamiltonian of which is given by, 
\begin{equation}
\label{H}
\hat{H}=\frac{1}{2m}\sum_{i=1}^{N}\hat{p}_i^2+\frac{K}{2}\sum_{i=1}^{N}\hat{q}_i^2+
\frac{k}{2}\sum_{i=1}^{N}(\hat{q}_{i+1}-\hat{q}_{i})^2.
\end{equation}   
Here   $\hat{q}_i,\ \hat{p}_i$ denote position, momentum 
operators of the oscillators, satisfying the canonical commutation relations 
$ [\hat{q}_i,\hat{p}_j]=i\hbar\, \delta_{i,j}$. It is convenient to define a 
$2N$-component operator column $\hat\xi$ of dimensionless variables, 
\begin{equation*}
\label{xi}
\hat\xi=\left(\begin{array}{c}\hat{Q}_i=\sqrt{\frac{m\omega}{\hbar}}\,q_i\\ 
\hat{P}_j=\sqrt{\frac{1}{m\omega\hbar}}\,p_j\end{array}\right),\ \  \omega^{2}=\frac{K}{m},\ i,j=1,2,\ldots, N,  
 \end{equation*} 
and express the commutation relations compactly as,  
\begin{eqnarray}
\label{com}
[\hat{\xi}_\alpha,\hat{\xi}_\beta]&=& i\, \Gamma _{\alpha\beta};\ \ \alpha,\beta=1,2,\ldots, 2N, \     
\end{eqnarray}
where, $\Gamma=\left(\begin{array}{cc}0 & I\\ -I & 0 \end{array}\right);$ 
$I$ denotes the $N\times N$ unit matrix. A general real homogeneous linear transformation on $\hat\xi$ 
preserving the canonical commutation relations (\ref{com}) is a $2N\times 2N$ symplectic 
transformation~\cite{NM} 
$S\in {\rm Sp}(2N, {\rm R})$ and there exists a corresponding unitary operator $\hat U(S)$ on the Hilbert space 
on which the operators $\hat\xi$ act:      
\begin{eqnarray*}
\hat U^\dag(S)\hat\xi_\alpha\hat U(S)
=\hat\xi'_\alpha=\sum_{\alpha'}S_{\alpha\alpha'}\hat\xi_{\alpha'}\nonumber \\  
{\rm \ such \ that\ } 
[\hat\xi'_\alpha,\hat\xi'_\beta]=i\, \Gamma _{\alpha\beta}\Rightarrow S\Gamma S^T=\Gamma.
\end{eqnarray*}

Here, we restrict to Gaussian quantum states of the system and the bath. These are completely characterized by 
the first and second moments of $\hat\xi,$ arranged conveniently in the form of $2N\times 2N$ covariance matrix 
$V$ as, 
\begin{equation*}
V_{\alpha\beta}=\frac{1}{2}\, \langle \{\bigtriangleup\hat\xi_{\alpha},\bigtriangleup\hat\xi_{\beta}\}\rangle,\ 
\ \alpha,\beta=1,2,\ldots , 2N,
\end{equation*}
 where $\bigtriangleup\hat\xi=\hat\xi-\langle\hat\xi\rangle$, $\{\hat O_1,\hat O_2\}=\hat O_1\hat O_2+\hat 
O_2\hat O_1$ 
 and $\langle \hat O\rangle~=~{\rm Tr}[\hat\rho \hat O]$ denotes the expectation value of the operator $\hat O$ 
in the quantum state $\hat\rho$. Under  symplectic transformation, a Gaussian state is mapped to another 
Gaussian state characterized by the covariance matrix $V'=SVS^T.$ 

Time evolution of the elements of the variance matrix under $\hat {\cal U}(t)={\rm exp}\{-it\hat H/\hbar\}$ may 
be identified as  a symplectic transformation (as the  Hamiltonian of Eq.~(\ref{H}) is a quadratic in the 
canonical operators): 
\begin{eqnarray*}
V_{\alpha\beta}(t)&=&\frac{1}{2}{\rm Tr}[\hat\rho(t)\, 
\{\bigtriangleup\hat\xi_{\alpha}(0),\bigtriangleup\hat\xi_{\beta}(0)\}]\nonumber \\
&=& \frac{1}{2}{\rm Tr}[\hat\rho(0)\, 
\hat {\cal U}^\dag(t)\{\bigtriangleup\hat\xi_{\alpha}(0),\bigtriangleup\hat\xi_{\beta}(0)\}\hat {\cal 
U}(t)]\nonumber \\
&=&\left({\cal S}(t)V(0){\cal S}^T(t)\right)_{\alpha\beta}
\end{eqnarray*}
or $V(t)={\cal S}(t)V(0){\cal S}^T(t)$, where ${\cal S}(t)$ denotes the $2N~\times~2N$ symplectic matrix 
corresponding to the unitary time evolution on the Hilbert space of the quantum state.  The explicit structure 
of the symplectic transformation matrix ${\cal S}(t)$ associated with the dynamical evolution 
$\hat {\cal U}(t)$ in the present model is readily identified, as will be outlined in the following.   
 
We first express the Hamiltonian (\ref{H}) in the following quadratic form  
\begin{equation}
\label{H2}
\hat H=  \frac{\omega\hbar}{2}\, \, \hat\xi^T \left( \begin{array}{cc} A & 0 \\ 0 & I \end{array}  
\right)\hat\xi
\end{equation}
where the elements of the $N\times N$ bolck matrix $A$ are given by, 
\begin{equation}
A_{i,j}=(2\epsilon+1)\, \delta_{i,j}-\epsilon\, \left[\delta_{i,j+1}+\delta_{i+1,j}\right],\ \ 
\epsilon=\frac{k}{K}.
\end{equation}
Identifying the  real orthogonal transformation $\sigma$ which diagonalizes the real symmetric matrix $A$ i.e., 
\begin{eqnarray}
\label{sigmalambda}
\sigma\, A \sigma^T&=&\Lambda={\rm diag}(\lambda(\phi_1),\lambda(\phi_2),\ldots, \lambda(\phi_N)),\nonumber \\
\lambda(\phi_l)&=& 1+2\epsilon\, (1-\cos\phi_l);\ \phi_l=\frac{l\pi}{N+1}
 \\
\sigma_{s,l}&=&\sqrt{\frac{2}{N+1}}\, \sin(s\phi_l),\nonumber  
\end{eqnarray} 
we express the Hamiltonian (\ref{H2}) in its  decoupled structure: 
\begin{eqnarray}
\label{hdiag}
\hat H&=&\frac{\omega\hbar}{2}\, \, \left({\cal S}\hat\xi\right)^{T} \left( \Lambda^{\frac{1}{2}} \oplus 
\Lambda^{\frac{1}{2}}\right)  \left({\cal S}\hat\xi\right) \nonumber \\
& =& \frac{\omega\hbar}{2}\,\hat U^\dag({\cal S})\left[ \sum_{l=1}^N\, \lambda_l^{\frac{1}{2}}\, \left( \hat 
P_l^{2}+ \hat Q_l^{2}\right)\right]\, \hat U({\cal S}), 
\end{eqnarray}
where  ${\cal S}=\Lambda^{\frac{1}{4}}\sigma \oplus \Lambda^{-\frac{1}{4}}\sigma$ 
is a symplectic transformation~\cite{fn} on the $2N$ component operator column  $\hat\xi$.

Thus, we obtain the $2N\times 2N$ symplectic matrix ${\cal S}(t)$ corresponding to the unitary time evolution 
operator $e^{-it\hat H/\hbar}$ as~\cite{fn2}:  
\begin{eqnarray}
\label{su}
{\cal S}(t)=\left(\begin{array}{cc}\cos(\omega t \, A^{\frac{1}{2}}) & A^{-\frac{1}{2}}\sin(\omega t \, 
A^{\frac{1}{2}})\\ 
-A^{\frac{1}{2}}\sin(\omega t \, A^{\frac{1}{2}})& \cos(\omega t \, A^{\frac{1}{2}}) \end{array}\right). 
\end{eqnarray} 
We proceed now to investigate the quantum evolution of a pure uncoupled squeezed state of oscillators.

\section{Time evolution of  pure uncoupled squeezed state} 
 
First, we decompose the global quantum state of $N$-oscillators into two parts:  $n<<N$ `system' oscillators 
i.e., a $n$ oscillator subsystem    and the rest of the whole state,  the bath.  
We consider an initial state of the whole system to be  pure product states of individual oscillators, 
\begin{eqnarray}
\label{inistate}
\vert\Psi(\eta,\mu,t=0)\rangle=\vert\phi_b(\eta,t=0)\rangle\otimes 
\vert\phi_b(\eta,t=0)\rangle\otimes\ldots\nonumber \\
\ \ \otimes \underbrace{\vert\chi_s(\mu,t=0)\rangle\otimes \ldots \otimes\vert\chi_s(\mu,t=0)\rangle}_{n {\rm\ 
system\ oscillators}}\nonumber \\ \otimes   
\vert\phi_b(\eta,t=0)\rangle\otimes \ldots \otimes\vert\phi_b(\eta,t=0)\rangle,\   
\end{eqnarray}
where the  oscillators in the bath  are  in the squeezed state 
\begin{equation}
\vert\phi_b(\eta,t=0)\rangle=\hat U(S(\eta))\vert 0\rangle.
\end{equation}
Here, $\hat U((S(\eta))$ denotes the squeezing operator (with associated $2\times 2$ symplectic matrix  given 
by~\cite{NM},
$S(\eta)~=~{\rm diag}(e^{-\eta/2},\ e^{\eta/2})$. 
The initial state of each of the system oscillators~\cite{fn3}, 
\begin{equation}
\vert\chi_s(\mu,t=0)\rangle=\hat U(S(\mu))\vert 0\rangle,
\end{equation}
is characterized by the squeezing parameter $\mu$.
(Here, $\vert 0\rangle$ denotes the ground state of the oscillator.) 

It may be noted that  initially, the whole system-bath state is a product state of oscillators and the 
subsystems  are not already in a stationary  state, when the couplings are switched on at $t=0^+$. In other 
words, the  global initial pure state (\ref{inistate}) is {\em  not} an energy eigenstate of the total 
Hamiltonian --  this being  a trivial case leaving the subsystems stationary under dynamical evolution. Also, in 
contrast to the case where the bath is 
initially in  thermal state  characterized by a temperature $T$, here the 
 bath is in a pure state specified by a squeezing parameter $\eta$.

The initial variance matrix of the system-bath pure state is given by, 
\begin{eqnarray}
\label{vxieta}
V(\eta,\mu; t=0)&=&\frac{1}{2}\left( D_Q(\eta,\mu)\oplus D_P(\eta,\mu)\right)
\end{eqnarray}
where the blocks $D_Q(\eta,\mu),$ and $D_P(\eta,\mu)$ are  $N\times N$ diagonal matrices,    
\begin{eqnarray}
\label{dx}
D_Q(\eta,\mu)&=&{\rm diag}(e^{-\eta}, \ldots , \underbrace{ e^{-\mu},\ldots, e^{-\mu}}_{r_1,r_2,\ldots r_n},\ 
\ldots , e^{-\eta})\nonumber \\
&=&D_P^{-1}(\eta,\mu).
\end{eqnarray} 

It is convenient to split the variance matrix $V(\eta, \mu; 0)$ as,   
\begin{eqnarray}
\label{vsplit}
V(\eta,\mu; 0)&=& v(\eta; 0)+\sum_{i=1}^{n}v_{i}(\eta,\mu; 0)\nonumber \\ 
{\rm where,}\ \  v(\eta; 0)&=&\frac{1}{2}\, \left(  e^{-\eta}\, I\oplus e^{\eta}\, I\right)\nonumber 
\\  
\ \ \ \ \ \ \ \sum_{i=1}^{n}v_{i}(\eta,\mu; 0)&=&V(\eta,\mu; 0)-v(\eta; 0).
\end{eqnarray}
 The non-zero elements of  
$v_{i}(\eta,\mu; 0)$ are readily identified as (see Eqs.~(\ref{vxieta}),(\ref{dx}), (\ref{vsplit})), 
\begin{eqnarray*}
\left[v_{i}(\eta,\mu; 0)\right]_{r_i,r_i}&=&\frac{1}{2}(e^{-\mu}-e^{-\eta}),\nonumber \\ 
\left[v_{i}(\eta,\mu; 0)\right]_{n+r_i,n+r_i}&=&\frac{1}{2}(e^{\mu}-e^{\eta})
\end{eqnarray*}

Temporal evolution of the quantum state (\ref{inistate}) is entirely determined by the 
symplectic transformation ${\cal S}(t)$ (given by Eq.~(\ref{su})) on the variance matrix, 
\begin{eqnarray}
\label{vblock}
V(\eta,\mu; t)&=&{\cal S}(t)V(\eta,\mu; 0){\cal S}^T(t)\nonumber \\
&=& v(\eta; t)+\sum_{i=1}^n\, v_{i}(\eta,\mu;t)\nonumber \\
&=& \left(\begin{array}{cc}V_{QQ}(\eta,\mu;t) & V_{QP}(\eta,\mu;t)\\ 
V^T_{QP}(\eta,\mu;t) & V_{PP}(\eta,\mu;t)\end{array} \right),
\end{eqnarray}
where $V_{QQ}(\eta,\mu;t), V_{PP}(\eta,\mu;t)$ and $V_{QP}(\eta,\mu;t)$ respectively denote the $N\times N$ 
diagonal and off-diagonal blocks of the variance matrix. 
 
In an infinitely long chain ($N\rightarrow \infty$),   closed form analytical expressions are obtained for the 
elements of the variance matrix $V(\eta,\mu;t)$  (by replacing the discrete variable `$\phi_l=\frac{l\pi}{N+1}$'   
of (\ref{sigmalambda}) by a continuous parameter `$\phi$\,' and   the sum `$\frac{1}{N+1}\sum_{l=1}^N$\,' by the 
integral `$\frac{1}{\pi}\int_0^\pi{\rm d}\phi$\, ' ):  
\begin{widetext}
\begin{eqnarray}
\label{csfun1}
\left[V_{QQ}(\eta,\mu;t)\right]_{s,l}&=&e^{-\eta}\, 
C^{(2,0)}_{s,l}(t)+e^{\eta}S^{(2,-1)}_{s,l}(t)+ 
 \frac{e^{-\mu}-e^{-\eta}}{2}\, \sum_{i=1}^{n}\, C^{(1,0)}_{s,r_i}(t)C^{(1,0)}_{l,r_i}(t)+
 \frac{e^{\mu}-e^{\eta}}{2}\, \sum_{i=1}^{n}\, 
S^{(1,-\frac{1}{2})}_{s,r_i}(t)S^{(1,-\frac{1}{2})}_{l,r_i}(t),\nonumber \\
\left[V_{PP}(\eta,\mu;t)\right]_{s,l}&=&e^{\eta}\, 
C^{(2,0)}_{s,l}(t)+e^{-\eta}S^{(2,1)}_{s,l}(t)+ 
 \frac{e^{\mu}-e^{\eta}}{2}\, \sum_{i=1}^{n}\, C^{(1,0)}_{s,r_i}(t)C^{(1,0)}_{l,r_i}(t) + 
\frac{e^{\mu}-e^{\eta}}{2}\, \sum_{i=1}^{n}\, 
S^{(1,\frac{1}{2})}_{s,r_i}(t)S^{(1,\frac{1}{2})}_{l,r_i}(t),\nonumber \\
\left[V_{QP}(\eta,\mu;t)\right]_{s,l}&=&-\frac{e^{-\eta}}{2}\, 
S^{(1,\frac{1}{2})}_{s,l}(2t)+\frac{e^{\eta}}{2}\, \, S^{(1,-\frac{1}{2})}_{s,l}(2t)-
 \frac{e^{-\mu}-e^{-\eta}}{2}\, \sum_{i=1}^{n}\, C^{(1,0)}_{s,r_i}(t)S^{(1,\frac{1}{2})}_{l,r_i}(t) \nonumber \\ 
&& + \frac{e^{-\mu}-e^{-\eta}}{2}\, \sum_{i=1}^{n}\, S^{(1,-\frac{1}{2})}_{s,r_i}(t)C^{(1,0)}_{l,r_i}(t),
\end{eqnarray}
where we have denoted, 
\begin{eqnarray}
\label{csfun2}
C^{(a,\kappa)}_{s,l}(t)&=&\frac{1}{\pi} \, \int_{0}^\pi\,{\rm d}\phi\,  
\sin(s\phi)\sin(l\phi)\, \lambda^{\kappa}(\phi)\, \cos^{a}[\omega t\lambda^{\frac{1}{2}}(\phi)] \nonumber \\
S^{(a,\kappa)}_{s,l}(t)&=&\frac{1}{\pi} \, \int_{0}^\pi\,{\rm d}\phi\,  
\sin(s\phi)\sin(l\phi)\, \lambda^{\kappa}(\phi)\sin^{a}[\omega t\lambda^{\frac{1}{2}}(\phi)] 
\end{eqnarray}
\end{widetext}
with $\lambda(\phi)=\epsilon\gamma^{-1}\, (1+2\gamma\cos\phi),\ \gamma=\frac{k}{K+2k}.$ 
These results are formally exact in the long chain limit. To make their meaning
evident, one resorts to  the weak coupling approximation $\epsilon\approx\gamma<<1,$  in which case the 
following standard form 
\begin{equation*}
\frac{1}{\pi}\int_{0}^\pi{\rm d}\phi \cos(s\phi)\cos[x(1-\gamma\cos\phi)]=J_{s}
(\gamma x)\cos(x-\frac{s\pi}{2})
\end{equation*}  
(where $J_s(x)$ denotes Bessel function of integral order) can be employed to simplify $C^{(a,\kappa)}_{s,l}(t), 
S^{(a,\kappa)}_{s,l}(t)$ of (\ref{csfun2}) up to order $O(\gamma)$ -- thus reducing  the  elements of the 
variance matrix $V(\eta,\mu;t)$ (given by Eq.~(\ref{csfun1}))  to time dependent   $\cos(\Omega t) {\rm \ or}\  
\sin(\Omega t)$ functions,   oscillating rapidly with Bessel functions $J_s(\gamma\Omega t)$  being the 
amplitudes (with $\Omega=\sqrt{\frac{K+2k}{m}}$). The long time  behavior of the system gets specified  by the  
asymptotic decay of Bessel functions i.e., $\displaystyle\lim_{x\rightarrow \infty}\, J_s(x)\rightarrow \vert 
x\vert^{-\frac{1}{2}}$  for $x>>s$. More specifically, we  find,   
\begin{eqnarray}
C^{(2,0)}_{s,l}&\rightarrow &\frac{1}{4}\, \delta_{s,l},\hskip 0.2in \ S^{(1,\pm \frac{1}{2})}_{s,l}, 
C^{(1,0)}_{s,l}\rightarrow 0, 
\nonumber \\  
S^{(2,\pm 1)}_{s,l}&\rightarrow&  
\frac{1}{4}\left(\delta_{s,l}\mp\gamma\left[\delta_{s,l+1}+\delta_{s,l-1}\right]\right), 
\end{eqnarray} 
in the limit $\gamma\Omega t\rightarrow \infty.$

It is thus evident that a subsystem of $n<<N$ oscillator   relaxes 
in a steady state, specified by the $2n\times 2n$ variance matrix, $V^{(n)}=V^{(n)}_{Q}\oplus V^{(n)}_{P},$ with 
elements,  
\begin{eqnarray}
\left[V^{(n)}_{Q}\right]_{sl}&=&   \frac{1}{2}\left(\cosh\eta+\frac{e^{-\eta}\gamma}{2}[\delta_{s,l+1}
+\delta_{s,l-1}]\right)\nonumber \\
\left[V^{(n)}_{P}\right]_{sl}&=&   \frac{1}{2}\left(\cosh\eta-\frac{e^{\eta}\gamma}{2}[\delta_{s,l+1}
+\delta_{s,l-1}]\right)
\end{eqnarray}    
(which exhibit a correlation $\langle \hat Q_{j}\hat Q_{j\pm 1}\rangle=\frac{\gamma}{2}\, e^{-\eta}$ and 
an anticorrelation $\langle \hat P_{j}\hat P_{j\pm 1}\rangle=-\frac{\gamma}{2}\, e^{\eta}$ between neighbors). 
Evidently, the equilibrium state of the `system' is independent of its initial form (i.e., it does not contain  
the  squeezing parameter $\mu$ of the initial system oscillators), but  depends on the squeezing parameter 
$\eta$ of the bath.

In particular, the variance matrix $V^{(1)}(t)$ of any single mode subsystem of the dynamically evolving global 
quantum state eventually converges, in the limit $ t>>(\gamma\Omega)^{-1}$, to an `equilibrium' structure, 
\begin{eqnarray}
V^{(1)}=\frac{1}{2}\, \left(\begin{array}{cc} 
   \cosh\eta & 0\\ 
   0 & \cosh\eta
   \end{array}\right),
\end{eqnarray}
which corresponds to a mixed density matrix of the familiar Boltzmann form, 
\begin{eqnarray}
\hat\rho^{(1)}&=&\frac{e^{-\beta\hbar\omega\hat H_{r}}}{{\rm Tr}[e^{-\beta\hbar\omega\hat H_{r}}]}\, ,\ \ 
\hat H_{r}= \frac{\hat{p}_r^2}{2m}+\frac{K}{2}\hat{q}_r^2\hskip 0.1in .   
\end{eqnarray} 
The inverse temperature $\beta$ 
is related to 
the squeezing parameter $\eta$ of the bath via, $\beta=\frac{2}{\hbar\omega}\, \coth^{-1}[\cosh\eta]$. 
 The  purity~\cite{Il} of the single oscillator equilibrium state $\nu^{(1)}~=~{\rm 
Tr}[(\rho^{(1)})^2]=[2\sqrt{\det(V^{(1)})}]^{-1}=[\cosh\eta]^{-1}<1$ captures the system-bath entanglement.    
This also reflects in the increase of the von Neumann entropy~\cite{Il} of the state from its initial value zero 
to the equilibrium value 
\begin{eqnarray*}
S(\hat\rho^{(1)})&=&-{\rm Tr}[\hat\rho^{(1)}\ln\hat\rho^{(1)}]\nonumber \\
&=& 
\left(\frac{1-\nu^{(1)}}{2\nu^{(1)}}\right)\ln\left(\frac{1+\nu^{(1)}}{1-\nu^{(1)}}\right)-\ln\left(\frac{2\nu^{
(1)}}{1+\nu^{(1)}}\right).
\end{eqnarray*}

Any two oscillator subsystem is found to eventually relax in a 'stationary' state specified  by the two mode 
variance matrix, 
\begin{equation}
V^{(2)}=\frac{1}{2}\left(\begin{array}{cccc} \cosh\eta &   \frac{e^{-\eta}\gamma}{2} & 0 & 0 \\
\frac{e^{-\eta}\gamma}{2} & \cosh\eta & 0 & 0 \\
0 & 0 & \cosh\eta &   -\frac{e^{\eta}\gamma}{2}\\
0 & 0& -\frac{e^{\eta}\gamma}{2} & \cosh\eta
\end{array}\right). 
\end{equation}

The stationary density matrix of the two oscillator system  has its purity, 
$\nu^{(2)}=[4\sqrt{\det(V^{(2)})}]^{-1}\approx [\cosh\eta]^{-2}$ 
which is clearly less than 1, and reveals the system-bath entanglement. 
One finds internal entanglement between the two oscillators if~\cite{Simon} 
$(\cosh^2\eta-\frac{e^{-2\eta}\gamma^2}{4})(\cosh^2\eta-\frac{e^{2\eta}\gamma^2}{4})-\cosh2\eta<0,$ 
and this happens when the squeezing parmeter $\eta$ of the bath is comparable in magnitude with the coupling 
$\gamma.$ Thus, one finds a tradeoff of entanglement within the  system oscillators and that between the 
system-bath -- as survival of internal entanglement implies nearly vanishing mixedness of the two oscillator 
system (i.e., $\nu^{(2)}\approx 1$ for small squeezing parameter).

\section{Conclusions}
In summary, we have shown, through an explicit analysis under the assumptions of continuum limit $N\rightarrow 
\infty$ and  weak coupling approximation $\epsilon\approx \gamma<< 1$, that any small subsystem of the whole 
pure squeezed state of $N$ oscillators evolving under hamonically coupled Hamiltonian, approaches equilibrium. 
This provides an excellent example in which stationary behaviour of any small subsystem of a large global pure 
state is shown to emerge as a natural phenomena consequent to  quantum dynamical evolution    - without the aid 
of any additional statistical postulates~\cite{SP}.  Equilibrated subsystems consisting of $1, 2, \ldots, n<<N$ 
oscillators are found to be mixed, revealing their quantum entanglement with the rest of the system i.e., the 
bath. A single mode oscillator  'system' is shown to relax in a mixed density matrix in the Boltzmann canonical 
form. The connection between the squeezing parameter of the bath and the temperature is a new feature of our 
work. We also find (in the case of $n=2$) that entanglement within the `system' oscillators in the steady state 
survives only when the squeezing parameter and the coupling strength are of comparable magnitude. These features 
on two and more subsystem oscillators, to the best of our knowledge, have not been recorded in the literature.         

The long chain limit  and the weak coupling approximation  made our analysis amenable to analytical results. One 
has to resort to numerical approach to evaluate the integrals (\ref{csfun2}) in the continuum, strong coupling 
limits. Also, the finite $N$ limit may be addressed with the help of numerical investigations both in the strong 
and weak coupling limits. These issues would be of interest from a foundational point of view and we plan to 
address this issue in a separate communication.

\section*{ACKNOWLEDGEMENTS}
\noindent ARU acknowledges financial support of the Commonwealth Commission, UK and thanks Professor Sandu 
Popescu and Paul Skrzypczyk for insightful comments.


\begin{thebibliography}{1}
\bibitem{fn0} The terminology `equilibrium' is used here in a rather non-technical sense, to imply stationary  
long time behaviour.
\bibitem{fn1} Even when the individual pure state of the whole system has zero entropy, subsystems are in a 
mixed state, exhibiting non-zero entropy, thus revealing 'objective' randomness. This is in stark  contrast with  
classical statistical mechanics, where the probability distributions that define entropy reflect our 
'subjective' lack of knowledge.
\bibitem{fn} The  $2N\times 2N$ matrix  ${\cal 
S}~=~\Lambda^{\frac{1}{4}}\sigma~\oplus~\Lambda^{-\frac{1}{4}}\sigma$ is a symplectic transformation belonging 
to  Sp$(2n$,R) as can be explicitly verified via the defining relation ${\cal S} \Gamma {\cal S}^T=\Gamma$.
\bibitem{fn2} 
The $2\times 2$ symplectic transformation  corresponding to the unitary operation $\hat 
U(S(\phi))=e^{\frac{-i\phi}{2}(\hat P^2+\hat Q^2)}$ on the single oscillator Hilbert space is given by~\cite{NM} 
$S(\phi)~=~\left(\begin{array}{cc} \cos\phi & \sin\phi \\ -\sin\phi & \cos\phi \end{array}\right).$ Employing 
this in Eq.~(\ref{hdiag}) for $N$-mode oscillators and simplifying yields ${\cal S}(t)$ of  Eq.~(\ref{su}).
\bibitem{fn3} The `system' could be isolated oscillators spread throughout the whole quantum state. It is for 
the sake of mathematical convenience that they are  grouped next to each other.
\end{thebibliography}
\end{document}